\documentclass[epjCONF, onecolumn]{svjour}
\usepackage{amsfonts}
\usepackage{amsmath}
\usepackage{graphicx}
\usepackage{subfigure}
\usepackage{dcolumn}
\usepackage{bm}
\usepackage[varg]{txfonts}
\usepackage[latin1]{inputenc}
\usepackage{booktabs}
\usepackage[dvips]{color}
\makeatletter
\newcommand{\figcaption}{\def\@captype{figure}\caption}
\newcommand{\tabcaption}{\def\@captype{table}\caption}

\newcommand{\Rmnum}[1]{\expandafter\@slowromancap\romannumeral #1@}
\def\hlinewd#1{%
  \noalign{\ifnum0=`}\fi\hrule \@height #1 \futurelet
   \reserved@a\@xhline}
\makeatother

\def\qq{\langle\bar qq\rangle}

\def\GGb{\langle g_s^2GG\rangle}
\def\qGq{\langle\bar qg_s\sigma Gq\rangle}

\def\f(s){[(\alpha+\beta)m_c^2-\alpha\beta s]}
\def\non{\\ \nonumber}
\session-title{Hadron Nuclear Physics (HNP) 2011}

\begin{document}

\title{Spin-1 charmonium-like states in QCD sum rule}

\author{Wei Chen\inst{1}\fnmsep\thanks{\email{boya@pku.edu.cn}}\and Shi-Lin Zhu\inst{1,2}\fnmsep\thanks{\email{zhusl@pku.edu.cn}}}

\institute{Department of Physics
and State Key Laboratory of Nuclear Physics and Technology,
Peking \\ University, Beijing 100871, China \and Technology
and Center of High Energy Physics, Peking University, Beijing
100871, China  }

\abstract{We study the possible spin-1 charmonium-like states by using
QCD sum rule approach. We calculate the two-point correlation
functions for all the local form tetraquark interpolating currents
with $J^{PC}=1^{--}, 1^{-+}, 1^{++}$ and $1^{+-}$ and extract the
masses of the tetraquark charmonium-like states. The mass of the
$1^{--}$ $qc\bar q\bar c$ state is $4.6\sim4.7$ GeV, which implies
a possible tetraquark interpretation for $Y(4660)$ meson.
The masses for both the $1^{++}$ $qc\bar q\bar c$ and $sc\bar s\bar c$
states are $4.0\sim 4.2$ GeV, which is slightly above the mass of $X(3872)$.
For the $1^{-+}$ and $1^{+-}$ $qc\bar q\bar c$ states, the extracted masses
are $4.5\sim4.7$ GeV and $4.0\sim 4.2$ GeV respectively.}

\maketitle

\section{Introduction}\label{sec:Introduction}
The underlying structures of the so-called $X, Y, Z$ states
are not understood well~\cite{2006-Swanson-p243-305,2008-Zhu-p283-322,2007-Rosner-p12002-12002}.
Some of them do not fit in the conventional
quark model easily and are considered as the candidates of the exotic
states beyond the quark model, such as the molecular states,
tetraquark states, the charmonium hybrid mesons, baryonium states
and so on. $X(3872)$ is the best studied charmonium-like state since
its discovery by the Belle
Collaboration~\cite{2003-Choi-p262001-262001}. Although the
analysis of angular distributions favors the assignment
$J^{PC}=1^{++}$~\cite{2005-Abe-p-,},
the $2^{-+}$ possibility is not ruled
out~\cite{2010-AmoSanchez-p11101-11101}. The mass and decay mode
of $X(3872)$ are very different from that of the $2^3P_1$ $c\bar
c$ state. Up to now, the possible interpretations of $X(3872)$
include the molecular state~\cite{2009-Liu-p411-428,2008-Liu-p63-73,2004-Swanson-p197-202,2004-Swanson-p189-195},
tetraquark state~\cite{2007-Matheus-p14005-14005,2007-Maiani-p182003-182003},
cusp~\cite{2004-Bugg-p8-14} and hybrid
charmonium~\cite{2003-Close-p210-216}. $Y(4260), Y(4360)$ and
$Y(4660)$ are the $Y$($J^{PC}=1^{--}$) family states discovered in the
initial state radiation (ISR) process. These new states lie above the
open charm threshold.
However, the $Y\rightarrow D^{(\ast)}\bar{D}^{(\ast)}$ decay modes
have not been observed yet~\cite{2007-Abe-p92001-92001}, which are
predicted to be the dominant decay modes of the charmonium above
the open charm threshold in the potential model. In
Refs.~\cite{2010-Wang-p323-332,2009-Albuquerque-p53-66}, the
authors studied the $1^{--}$ charmonium-like Y mesons using the
QCD sum rule approach. Maiani \textit{et al.} tried to assign
$Y(4260)$ as the $sc\bar s\bar c$ tetraquark in a P-wave
state~\cite{2005-Maiani-p031502-031502}. $Y(4260)$ was also
interpreted as the interesting charmonium hybrid
state~\cite{2005-Zhu-p212-212,2005-Close-p215-222}.
$Y(4660)$ was considered as a
$\psi(2S)f_0(980)$ bound state in Ref.~\cite{2008-Guo-p26-29}.
Recently there have been some efforts on the $1^{-+}$ charmonium-like
exotic states. For example, the structure of $X(4350)$ was studied using a
$D^{\ast}_sD_{s0}^{\ast}$ current with $J^{PC}=1^{-+}$~\cite{2010-Albuquerque-p-}.
Moreover, the newly observed state $Y(4140)$ was argued as a $1^{-+}$ exotic
charmonium hybrid state~\cite{2009-Mahajan-p228-228}.

In this contribution, we would like to report the systematic study of
the tetraquark charmonium-like states with $J^{PC}=1^{--}, 1^{-+}, 1^{++}$
and $1^{+-}$. After constructing all the tetraquark interpolating currents with
definite quantum numbers, we investigate the two-point correlation functions and
extract the masses of the charmonium-like states with QCD sum rule. We study both
the $qQ\bar q\bar Q$ and $sQ\bar s\bar Q$ systems where $Q=c, b$. We also discuss
the possible decay modes and experimental search of the charmonium-like states.
%
%
\section{INDEPENDENT CURRENTS}\label{curr}
Firstly, we construct all the local form diquark-antidiquark type of interpolating
currents in a systematic way. By considering the Lorentz structures, the charge
conjugation properties and the color symmetries of the tetraquark operators, we arrive
the following tetraquark interpolating currents with
$J^{PC}=1^{-+}, 1^{--}, 1^{++}$ and $1^{+-}$:
\begin{itemize}
\item The interpolating currents with $J^{PC}=1^{-+}$ and $1^{--}$
are:
\begin{eqnarray}
\nonumber J_{1\mu}&=&q_{a}^TC\gamma_5c_{b}(\bar{q}_{a}\gamma_{\mu}\gamma_5C\bar{c}^T_{b}+\bar{q}_{b}\gamma_{\mu}\gamma_5C\bar{c}^T_{a})
\pm
q_{a}^TC\gamma_{\mu}\gamma_5c_{b}(\bar{q}_{a}\gamma_5C\bar{c}^T_{b}+\bar{q}_{b}\gamma_5C\bar{c}^T_{a})\,
, \non J_{2\mu}&=&q_{a}^TC\gamma^{\nu}c_{b}(\bar{q}_{a}\sigma_{\mu\nu}C\bar{c}^T_{b}-\bar{q}_{b}\sigma_{\mu\nu}C\bar{c}^T_{a})
\pm
q_{a}^TC\sigma_{\mu\nu}c_{b}(\bar{q}_{a}\gamma^{\nu}C\bar{c}^T_{b}-\bar{q}_{b}\gamma^{\nu}C\bar{c}^T_{a})\,
, \non J_{3\mu}&=&q_{a}^TC\gamma_5c_{b}(\bar{q}_{a}\gamma_{\mu}\gamma_5C\bar{c}^T_{b}-\bar{q}_{b}\gamma_{\mu}\gamma_5C\bar{c}^T_{a})
\pm
q_{a}^TC\gamma_{\mu}\gamma_5c_{b}(\bar{q}_{a}\gamma_5C\bar{c}^T_{b}-\bar{q}_{b}\gamma_5C\bar{c}^T_{a})\,
, \non J_{4\mu}&=&q_{a}^TC\gamma^{\nu}c_{b}(\bar{q}_{a}\sigma_{\mu\nu}C\bar{c}^T_{b}+\bar{q}_{b}\sigma_{\mu\nu}C\bar{c}^T_{a})
\pm
q_{a}^TC\sigma_{\mu\nu}c_{b}(\bar{q}_{a}\gamma^{\nu}C\bar{c}^T_{b}+\bar{q}_{b}\gamma^{\nu}C\bar{c}^T_{a})\,
,
\\ \label{currents1}
J_{5\mu}&=&q_{a}^TCc_{b}(\bar{q}_{a}\gamma_{\mu}C\bar{c}^T_{b}+\bar{q}_{b}\gamma_{\mu}C\bar{c}^T_{a})
\pm
q_{a}^TC\gamma_{\mu}c_{b}(\bar{q}_{a}C\bar{c}^T_{b}+\bar{q}_{b}C\bar{c}^T_{a})\,
, \non J_{6\mu}&=&q_{a}^TC\gamma^{\nu}\gamma_5c_{b}(\bar{q}_{a}\sigma_{\mu\nu}\gamma_5C\bar{c}^T_{b}+\bar{q}_{b}\sigma_{\mu\nu}\gamma_5C\bar{c}^T_{a})
\pm
q_{a}^TC\sigma_{\mu\nu}\gamma_5c_{b}(\bar{q}_{a}\gamma^{\nu}\gamma_5C\bar{c}^T_{b}+\bar{q}_{b}\gamma^{\nu}
\gamma_5C\bar{c}^T_{a})\, , \non
J_{7\mu}&=&q_{a}^TCc_{b}(\bar{q}_{a}\gamma_{\mu}C\bar{c}^T_{b}-\bar{q}_{b}\gamma_{\mu}C\bar{c}^T_{a})
\pm
q_{a}^TC\gamma_{\mu}c_{b}(\bar{q}_{a}C\bar{c}^T_{b}-\bar{q}_{b}C\bar{c}^T_{a})\,
, \non
J_{8\mu}&=&q_{a}^TC\gamma^{\nu}\gamma_5c_{b}(\bar{q}_{a}\sigma_{\mu\nu}\gamma_5C\bar{c}^T_{b}-\bar{q}_{b}\sigma_{\mu\nu}\gamma_5C\bar{c}^T_{a})
\pm
q_{a}^TC\sigma_{\mu\nu}\gamma_5c_{b}(\bar{q}_{a}\gamma^{\nu}\gamma_5C\bar{c}^T_{b}-\bar{q}_{b}\gamma^{\nu}
\gamma_5C\bar{c}^T_{a})\, .
\end{eqnarray}
where ``$+$'' corresponds to $J^{PC}=1^{-+}$,
``$-$'' corresponds to $J^{PC}=1^{--}$.

\item The interpolating currents with $J^{PC}=1^{++}$ and $1^{+-}$
are:
\begin{eqnarray}
\nonumber J_{1\mu}&=&q^T_aCc_b(\bar{q}_a\gamma_{\mu}\gamma_5C\bar{c}^T_b+\bar{q}_b\gamma_{\mu}\gamma_5C\bar{c}^T_a)
\pm
q^T_aC\gamma_{\mu}\gamma_5c_b(\bar{q}_aC\bar{c}^T_b+\bar{q}_bC\bar{c}^T_a)\,
, \non
J_{2\mu}&=&q^T_aCc_b(\bar{q}_a\gamma_{\mu}\gamma_5C\bar{c}^T_b-\bar{q}_b\gamma_{\mu}\gamma_5C\bar{c}^T_a)
\pm
q^T_aC\gamma_{\mu}\gamma_5c_b(\bar{q}_aC\bar{c}^T_b-\bar{q}_bC\bar{c}^T_a)\,
, \non
J_{3\mu}&=&q^T_aC\gamma_5c_b(\bar{q}_a\gamma_{\mu}C\bar{c}^T_b+\bar{q}_b\gamma_{\mu}C\bar{c}^T_a)
\pm
q^T_aC\gamma_{\mu}c_b(\bar{q}_a\gamma_5C\bar{c}^T_b+\bar{q}_b\gamma_5C\bar{c}^T_a)\,
,
\\ \label{currents2}
J_{4\mu}&=&q^T_aC\gamma_5c_b(\bar{q}_a\gamma_{\mu}C\bar{c}^T_b-\bar{q}_b\gamma_{\mu}C\bar{c}^T_a)
\pm
q^T_aC\gamma_{\mu}c_b(\bar{q}_a\gamma_5C\bar{c}^T_b-\bar{q}_b\gamma_5C\bar{c}^T_a)\,
, \non
J_{5\mu}&=&q^T_aC\gamma^{\nu}c_b(\bar{q}_a\sigma_{\mu\nu}\gamma_5C\bar{c}^T_b+\bar{q}_b\sigma_{\mu\nu}\gamma_5C\bar{c}^T_a)
\pm
q^T_aC\sigma_{\mu\nu}\gamma_5c_b(\bar{q}_a\gamma^{\nu}C\bar{c}^T_b+\bar{q}_b\gamma^{\nu}C\bar{c}^T_a)\,
, \non
J_{6\mu}&=&q^T_aC\gamma^{\nu}c_b(\bar{q}_a\sigma_{\mu\nu}\gamma_5C\bar{c}^T_b-\bar{q}_b\sigma_{\mu\nu}\gamma_5C\bar{c}^T_a)
\pm
q^T_aC\sigma_{\mu\nu}\gamma_5c_b(\bar{q}_a\gamma^{\nu}C\bar{c}^T_b-\bar{q}_b\gamma^{\nu}C\bar{c}^T_a)\,
, \non
J_{7\mu}&=&q^T_aC\gamma^{\nu}\gamma_5c_b(\bar{q}_a\sigma_{\mu\nu}C\bar{c}^T_b+\bar{q}_b\sigma_{\mu\nu}C\bar{c}^T_a)
\pm
q^T_aC\sigma_{\mu\nu}c_b(\bar{q}_a\gamma^{\nu}\gamma_5C\bar{c}^T_b+\bar{q}_b\gamma^{\nu}\gamma_5C\bar{c}^T_a)\,
, \non
J_{8\mu}&=&q^T_aC\gamma^{\nu}\gamma_5c_b(\bar{q}_a\sigma_{\mu\nu}C\bar{c}^T_b-\bar{q}_b\sigma_{\mu\nu}C\bar{c}^T_a)
\pm
q^T_aC\sigma_{\mu\nu}c_b(\bar{q}_a\gamma^{\nu}\gamma_5C\bar{c}^T_b-\bar{q}_b\gamma^{\nu}\gamma_5C\bar{c}^T_a)\,
.
\end{eqnarray}
where ``$+$'' corresponds to $J^{PC}=1^{++}$, ``$-$'' corresponds to $J^{PC}=1^{+-}$.
\end{itemize}
The subscripts $a$ and $b$ are the color indices, $q$ denotes $u$ or $d$ quark. It is understood that all the currents
in Eqs.~(\ref{currents1})-(\ref{currents2}) should contain $(uc\bar u \bar c + dc\bar d \bar c)$ in order to have
definite isospin and $G$-parity. The details about the current construction could be found in Ref.~\cite{2011-Chen-p34010-34010}.

%
%
\section{SPECTRAL DENSITY}\label{sec:QSR}
In the past several decades, QCD sum rule has been widely used to
study the hadron structures and proven to be a very powerful
non-perturbative method~\cite{1979-Shifman-p385-447,1985-Reinders-p1-1}.
We consider the two-point
correlation function:
\begin{eqnarray}
\nonumber \Pi_{\mu\nu}(q^{2})&=& i\int
d^4xe^{iqx}\langle0|T[J_{\mu}(x)J_{\nu}^{\dag}(0)]|0\rangle
\\
&=&-\Pi_1(q^2)(g_{\mu\nu}-\frac{q_{\mu}q_{\nu}}{q^2})+\Pi_0(q^2)\frac{q_{\mu}q_{\nu}}{q^2},\label{equ:Pi}
\end{eqnarray}
where $J_{\mu}$ is a interpolating current for the tetraquark states.
$\Pi_1(q^2)$ is related to the vector meson while $\Pi_0(q^2)$ is
the scalar current polarization function. The correlation function
$\Pi_{\mu\nu}(q^{2})$ can be calculated in the operator product
expansion (OPE) using perturbative QCD augmented with non-perturbative
quark and gluon condensates to describe the large distance physics.
At the hadron level, the correlation function is expressed by the
dispersion relation with a spectral function:
\begin{eqnarray}
\Pi(q^2)=\int_{4m_c^2}^{\infty}\frac{\rho(s)}{s-q^2-i\epsilon},
\label{Phenpi}
\end{eqnarray}
In approximation of the infinitely narrow widths of resonances, the
spectral function can be expressed as:
\begin{eqnarray}
\nonumber
\rho(s)&\equiv&\sum_n\delta(s-m_n^2)\langle0|\eta|n\rangle\langle n|\eta^+|0\rangle\\
&=&f_X^2\delta(s-m_X^2)+...,   \label{Phenrho}
\end{eqnarray}
where ``...'' represents the higher states contribution.

The theoretical basis of the QCD sum rule approach is the assumption of
the quark-hadron duality, which ensures the equivalence of the
correlation functions obtained at the hadron level and the
quark-gluon level. After performing the Borel transformation to the
correlation functions, we can extract the mass of the state $X$:
\begin{eqnarray}
m_X^2=\frac{\int_{4m_c^2}^{s_0}dse^{-s/M_B^2}s\rho(s)}{\int_{4m_c^2}^{s_0}dse^{-s/M_B^2}\rho(s)}.
\label{mass}
\end{eqnarray}
where $s_0$ is the continuum threshold and $M_B$ is the Borel parameter.
We performed the QCD sum rule analysis for all the tetraquark currents in
Eqs.~(\ref{currents1})-(\ref{currents2}). The results of OPE can be
found in Ref.~\cite{2011-Chen-p34010-34010}.
\section{QCD Sum Rule Analysis}\label{sec:analysis}

We use the following parameter values of the quark masses and various
condensates~\cite{2010-Nakamura-p75021-75021,2001-Eidemuller-p203-210,1999-Jamin-p300-303}
for the numerical analysis:
$m_q(2\text{GeV})=(4.0\pm0.7)\text{ MeV}, m_s(2\text{GeV})=(101^{+29}_{-21})\text{MeV},
m_c(m_c)=(1.23\pm0.09)\text{GeV},
m_b(m_b)=(4.20\pm0.07)\text{GeV},
\qq=-(0.23\pm0.03)^3\text{GeV}^3, \qGq=-M_0^2\qq,
M_0^2=(0.8\pm0.2)\text{GeV}^2,
\langle\bar ss\rangle/\qq=0.8\pm0.2, \GGb=0.88\text{GeV}^4$.
There are two important parameters in
QCD sum rule analysis: the threshold parameter $s_0$ and the Borel
mass $M_B$. The stability of QCD sum rule requires a suitable working
region of $s_0$ and $M_B$. Since the exponential weight function in
Eq.~\ref{mass}, the higher state contribution is naturally
suppressed for small value of $M_B$. However, the OPE convergence would
become worse if $M_B$ was too small.
In our analysis, we choose the value of $s_0$ around which the variation
of the extracted mass $m_X$ with $M_B^2$ is minimum. The working
region of the Borel mass is determined by the convergence of the
OPE series and the pole contribution. The requirement of the
convergence of the OPE series leads to the lower bound $M^2_{min}$
of the Borel parameter while the constraint of the pole
contribution yields the upper bound of $M_B^2$.
\subsection{Vector charmonium-like systems}
For the interpolating currents with $J^{PC}=1^{-+}$ and
$1^{--}$, we keep the $m_q$ and $m_s$ related terms in the spectral
densities. These terms give important corrections to the OPE
series and are useful to enhance the stability of the sum
rule. For the $qc\bar q\bar c$ systems, the absolute value of the four quark
condensate $\qq^2$ is bigger than other condensates in the region of
$M_B^2<3.1$ GeV$^2$. It is the dominant power contribution to
the correlation function in this region.
Especially for the currents with $J^{PC}=1^{--}$, the quark
condensate $\qq$ is proportional to the light quark mass $m_q$ and
vanishes if we take $m_q=0$. However, it is proportional to the
strange quark mass $m_s$ and larger than $\langle\bar ss\rangle^2$ for the
$sc\bar s\bar c$ system since $m_s\gg m_q$. This is the main difference
between the $qc\bar q\bar c$ and $sc\bar s\bar c$ systems. The similar situation
exits in $1^{-+}$ charmonium-like systems.

After careful study of the OPE convergence and the pole contribution, we find the
suitable working region of the Borel parameter for each vector charmonium-like current.
The threshold value of $s_0$ is also fixed around which the variation
of the extracted mass $m_X$ with $M_B^2$ is minimum. In Fig.~\ref{fig1} and Fig.~\ref{fig2},
we show the variation of $m_X$ with
the threshold value $s_0$ and Borel parameter $M^2_B$ for the
current $J_{1\mu}$ with $J^{PC}=1^{--}$ in $qc\bar q\bar c$ and
$sc\bar s\bar c$ systems, respectively. One notes that they are very
similar with each other except the chosen $s_0$ mentioned above. The extracted
mass of the $qc\bar q\bar c$ state is $4.64$ GeV,
which is consistent with the mass of the meson $Y(4660)$. One may
wonder whether $Y(4660)$ could be a tetraquark state.
The extracted mass of the $sc\bar s\bar c$ state is $4.92$ GeV, which is about $0.28$ GeV
higher than that of the $qc\bar q\bar c$ state.
\begin{center}
\begin{tabular}{lr}
\scalebox{0.56}{\includegraphics{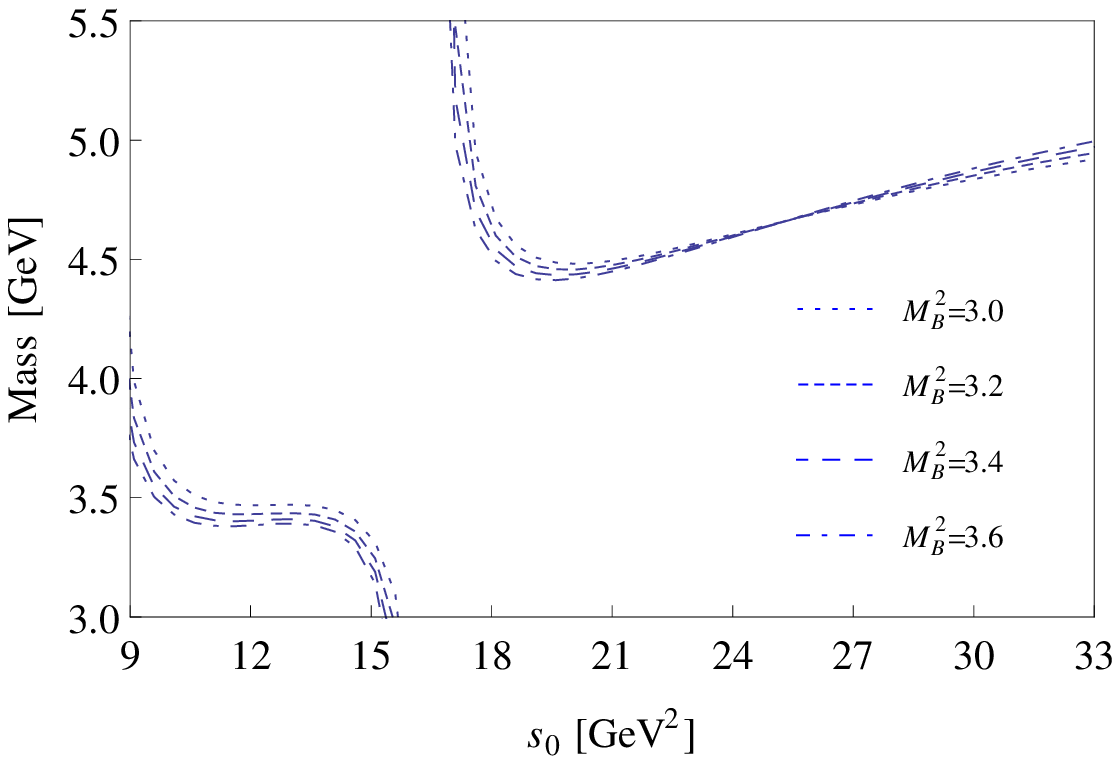}}&
\scalebox{0.56}{\includegraphics{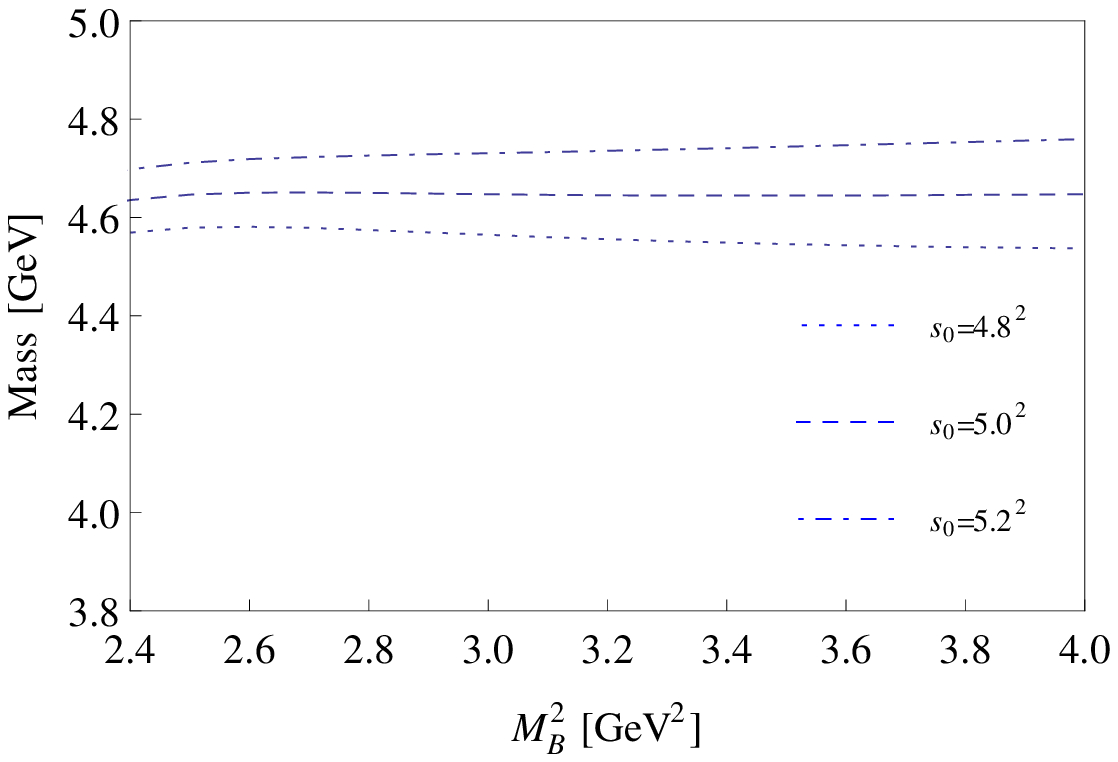}}
\end{tabular}
\centerline{\hspace{0.00in} {(a)} \hspace{3in}{ (b)}}
\figcaption{The variation of $m_X$ with $s_0$(a) and $M^2_B$(b)
corresponding to the current $J_{1\mu}$ for the $1^{--}$ $qc\bar
q\bar c$ system.} \label{fig1}
\end{center}
\begin{center}
\begin{tabular}{lr}
\scalebox{0.56}{\includegraphics{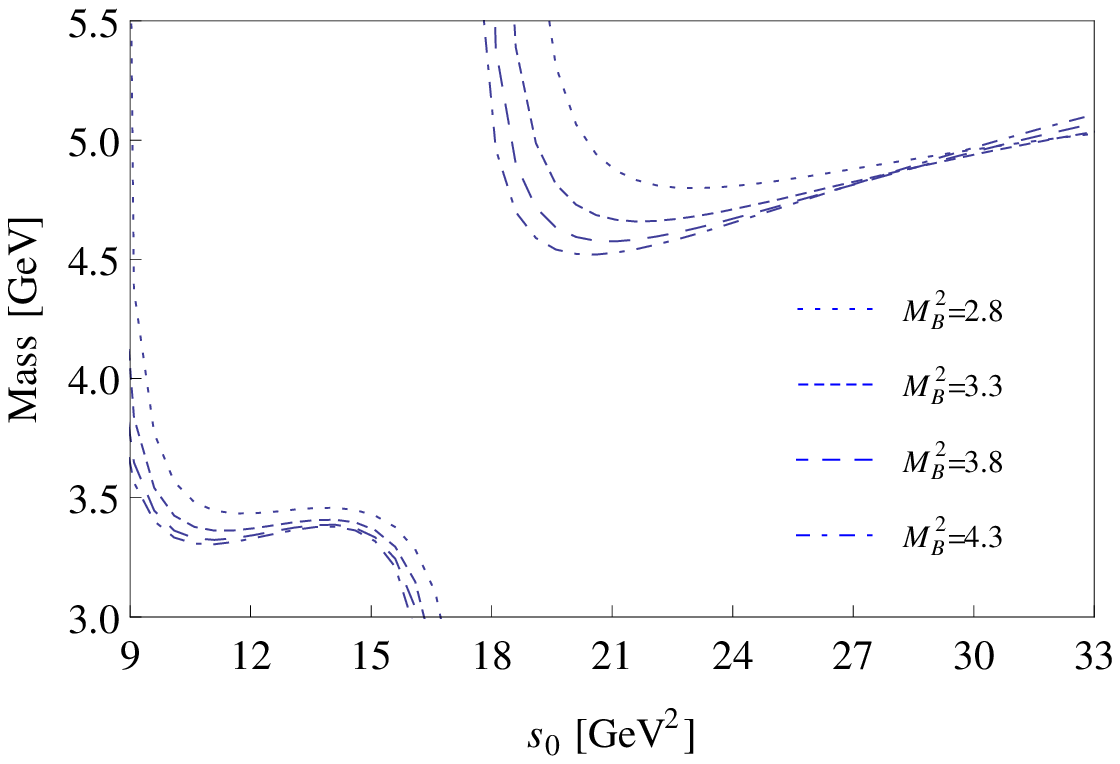}}&
\scalebox{0.56}{\includegraphics{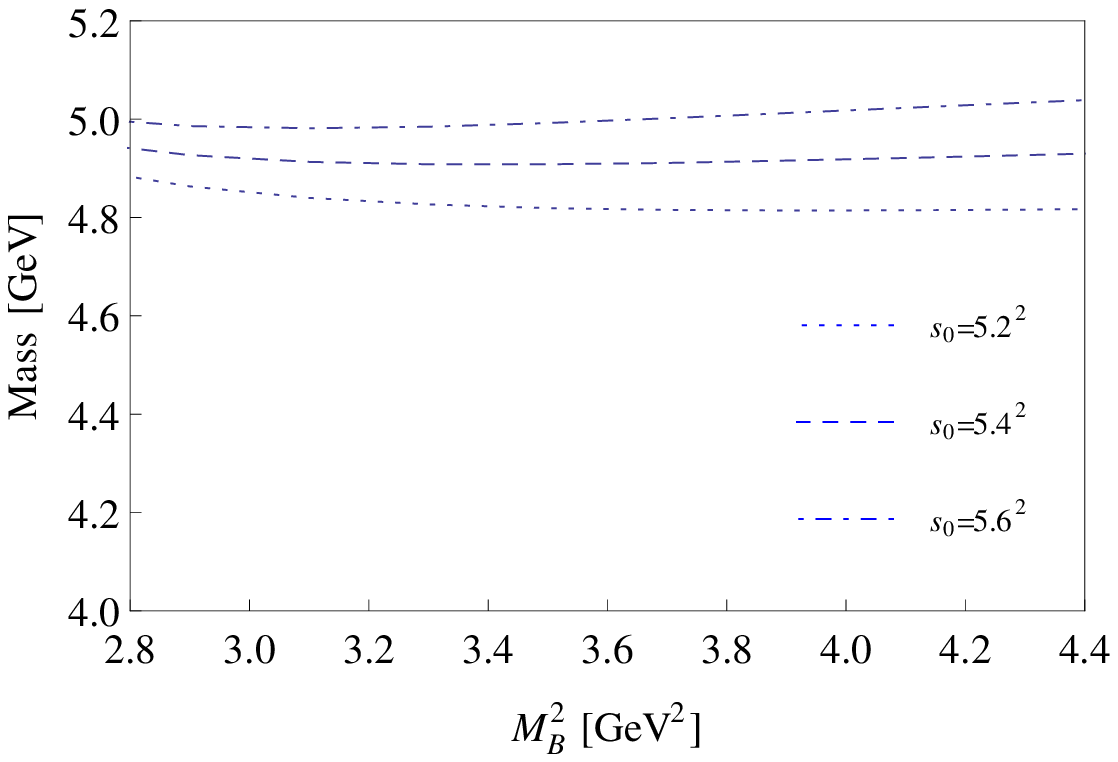}}
\end{tabular}
\centerline{\hspace{0.05in} {(a)} \hspace{3in}{ (b)}}
\figcaption{The variation of $m_X$ with $s_0$(a) and $M^2_B$(b)
corresponding to the current $J_{1\mu}$ for the $1^{--}$ $sc\bar
s\bar c$ system.} \label{fig2}
\end{center}

Performing the QCD sum rule analysis, we show the Borel window,
the threshold value, the extracted mass and the pole contribution
corresponding to the tetraquark currents with $J^{PC}=1^{--}$
in Table~\ref{table2}. The results of the $1^{-+}$ system
are listed in Table~\ref{table3}. We only present the numerical results
for the currents which lead to the stable mass sum rules in the
working range of the Borel parameter. For example, only the currents
$J_{1\mu}, J_{4\mu}$ and $J_{7\mu}$ with $J^{PC}=1^{--}$ in $qc\bar q\bar c$
systems have the reliable mass sum rules. For
$J_{2\mu}, J_{3\mu}, J_{5\mu}, J_{6\mu}$ and $J_{8\mu}$, the
stability is so bad that the extracted mass $m_X$ grows
monotonically with the threshold value $s_0$ and the Borel
parameter $M_B$. These currents may couple to the $1^{--}$ states
very weakly, leading to the above unstable mass sum rules.
We also study the bottomonium-like analogues by replacing $m_c$ with
$m_b$ in the correlation functions and repeating the same analysis
procedures done above. The numerical results of the $qb\bar
q\bar b$ and $sb\bar s\bar b$ systems are collected in Table~\ref{table2} and
Table~\ref{table3}.
\begin{center}
\begin{tabular}{cccccc}
\hlinewd{.8pt}
                   & Currents & $s_0(\mbox{GeV}^2)$&$[M^2_{\mbox{min}}$,$M^2_{\mbox{max}}](\mbox{GeV}^2)$&$m_X$\mbox{(GeV)}&PC(\%)\\
\hline
                        & $J_{1\mu}$      &  $5.0^2$         & $2.9\sim3.6$           & $4.64\pm0.09$     & 44.1  \\
$qc\bar q\bar c$ system & $J_{4\mu}$      &  $5.0^2$         & $2.9\sim3.6$           & $4.61\pm0.10$     & 46.4  \\
                        & $J_{7\mu}$      &  $5.2^2$         & $2.9\sim4.1$           & $4.74\pm0.10$     & 47.3
\vspace{5pt} \\
                        & $J_{1\mu}$      &  $5.4^2$         & $2.8\sim4.5$           & $4.92\pm0.10$     & 50.3  \\
                        & $J_{2\mu}$      &  $5.0^2$         & $2.8\sim3.5$           & $4.64\pm0.09$     & 48.6  \\
$sc\bar s\bar c$ system & $J_{3\mu}$      &  $4.9^2$         & $2.8\sim3.4$           & $4.52\pm0.10$     & 45.6  \\
                        & $J_{4\mu}$      &  $5.4^2$         & $2.8\sim4.5$           & $4.88\pm0.10$     & 51.7  \\
                        & $J_{7\mu}$      &  $5.3^2$         & $2.8\sim4.3$           & $4.86\pm0.10$     & 46.0  \\
                        & $J_{8\mu}$      &  $4.8^2$         & $2.8\sim3.1$           & $4.48\pm0.10$     & 43.2
\vspace{5pt} \\
$qb\bar q\bar b$ system & $J_{7\mu}$      &  $11.0^2$        &
$7.2\sim8.5$           & $10.51\pm0.10$    & 45.8
\vspace{5pt} \\
                        & $J_{1\mu}$      &  $11.0^2$        & $7.2\sim8.3$           & $10.60\pm0.10$    & 47.0  \\
                        & $J_{2\mu}$      &  $11.0^2$        & $7.2\sim8.4$           & $10.55\pm0.11$    & 43.6  \\
$sb\bar s\bar b$ system & $J_{3\mu}$      &  $11.0^2$        & $7.2\sim8.4$           & $10.55\pm0.10$    & 43.7  \\
                        & $J_{4\mu}$      &  $11.0^2$        & $7.2\sim8.4$           & $10.53\pm0.11$    & 44.3  \\
                        & $J_{7\mu}$      &  $11.0^2$        & $7.2\sim8.2$           & $10.62\pm0.10$    & 42.0  \\
                        & $J_{8\mu}$      &  $11.0^2$        & $7.2\sim8.4$           & $10.53\pm0.10$    & 44.1  \\
\hlinewd{.8pt}
\end{tabular}
\tabcaption{The threshold value, Borel window, mass and pole
contribution corresponding to the currents with $J^{PC}=1^{--}$ in the
$qc\bar q\bar c$, $sc\bar s\bar c$, $qb\bar q\bar b$ and $sb\bar
s\bar b$ systems.\label{table2}}
\end{center}
\begin{center}
\begin{tabular}{cccccc}
\hlinewd{.8pt}
                   & Currents & $s_0(\mbox{GeV}^2)$&$[M^2_{\mbox{min}}$,$M^2_{\mbox{max}}](\mbox{GeV}^2)$&$m_X$\mbox{(GeV)}&PC(\%)\\
\hline
                        & $J_{6\mu}$      &  $5.1^2$             & $2.9\sim3.9$              & $4.67\pm0.10$    & 50.2  \\
$qc\bar q\bar c$ system & $J_{7\mu}$      &  $5.2^2$             & $2.9\sim4.2$              & $4.77\pm0.10$    & 47.4  \\
                        & $J_{8\mu}$      &  $4.9^2$             & $2.9\sim3.4$              & $4.53\pm0.10$    & 46.3
\vspace{5pt}\\
                        & $J_{1\mu}$      &  $5.0^2$             & $2.9\sim3.4$              & $4.67\pm0.10$    & 44.3  \\
                        & $J_{2\mu}$      &  $5.0^2$             & $2.9\sim3.4$              & $4.65\pm0.09$    & 45.6  \\
                        & $J_{3\mu}$      &  $4.9^2$             & $2.9\sim3.3$              & $4.54\pm0.10$    & 44.4  \\
                        & $J_{4\mu}$      &  $5.1^2$             & $2.9\sim3.7$              & $4.72\pm0.09$    & 44.8  \\
$sc\bar s\bar c$ system & $J_{5\mu}$      &  $5.0^2$             & $2.9\sim3.6$              & $4.62\pm0.10$    & 42.8  \\
                        & $J_{6\mu}$      &  $5.3^2$             & $2.9\sim4.3$              & $4.84\pm0.10$    & 47.3  \\
                        & $J_{7\mu}$      &  $5.3^2$             & $2.9\sim4.3$              & $4.87\pm0.10$    & 46.2  \\
                        & $J_{8\mu}$      &  $5.2^2$             & $2.9\sim4.1$              & $4.77\pm0.10$    & 44.1
\vspace{5pt}\\
                        & $J_{6\mu}$      &  $11.0^2$            & $7.2\sim8.6$              & $10.53\pm0.11$   & 44.2  \\
$qb\bar q\bar b$ system & $J_{7\mu}$      &  $11.0^2$            & $7.2\sim8.6$              & $10.53\pm0.10$   & 44.1  \\
                        & $J_{8\mu}$      &  $11.0^2$            & $7.2\sim8.6$              & $10.49\pm0.11$   & 44.7
\vspace{5pt}\\
                        & $J_{4\mu}$      &  $11.0^2$            & $7.2\sim8.1$              & $10.62\pm0.10$   & 41.2  \\
                        & $J_{5\mu}$      &  $11.0^2$            & $7.2\sim8.4$              & $10.56\pm0.10$   & 43.8  \\
$qb\bar q\bar b$ system & $J_{6\mu}$      &  $11.0^2$            & $7.2\sim8.3$              & $10.63\pm0.10$   & 42.4  \\
                        & $J_{7\mu}$      &  $11.0^2$            & $7.2\sim8.3$              & $10.62\pm0.09$   & 42.5  \\
                        & $J_{8\mu}$      &  $11.0^2$            & $7.2\sim8.3$              & $10.59\pm0.10$   & 43.1  \\
\hlinewd{.8pt}
\end{tabular}
\tabcaption{The threshold value, Borel window, mass and pole
contribution corresponding to the currents with $J^{PC}=1^{-+}$ in the
$qc\bar q\bar c$, $sc\bar s\bar c$, $qb\bar q\bar b$ and $sb\bar
s\bar b$ systems.\label{table3}}
\end{center}

\subsection{Axial-vector charmonium-like systems}
In this channel, the QCD sum rule analysis shows that the quark condensate $\qq$ is
the dominant correction to the correlation function for all the currents.
Although the OPE convergence becomes worse than that in the vector channel,
the currents $J_{5\mu}, J_{6\mu}, J_{7\mu}, J_{8\mu}$ have better OPE convergence
than that of $J_{1\mu}, J_{2\mu}, J_{3\mu}, J_{4\mu}$.
For the interpolating currents $J_{3\mu}$ and $J_{4\mu}$ with $J^{PC}=1^{++}$, we
obtain the working region of the Borel parameter in $3.0\leq
M_B^2\leq3.4$ GeV$^2$ for both the $qc\bar q\bar c$ and $sc\bar s\bar c$ systems.
The extracted mass is about $m_X=4.0\sim4.2$ GeV, which is slightly above
the mass of $X(3872)$. The $qb\bar q\bar b$ and $sb\bar s\bar b$
systems can be studied conveniently by replacement of the
parameters, including the quark masses and the various
condensates. The numerical results are listed in
Table~\ref{table4} for the $1^{++}$ systems and Table~\ref{table5}
for the $1^{+-}$ systems.
\begin{center}
\begin{tabular}{cccccc}
\hlinewd{.8pt}
                   & Currents & $s_0(\mbox{GeV}^2)$&$[M^2_{\mbox{min}}$,$M^2_{\mbox{max}}](\mbox{GeV}^2)$&$m_X$\mbox{(GeV)}&PC(\%)\\
\hline
$qc\bar q\bar c$ system & $J_{3\mu}$         &  $4.6^2$         & $3.0\sim3.4$           & $4.19\pm0.10$     & 47.3 \\
                        & $J_{4\mu}$         &  $4.5^2$         & $3.0\sim3.3$           & $4.03\pm0.11$     & 46.8
\vspace{5pt} \\
$sc\bar s\bar c$ system & $J_{3\mu}$         &  $4.6^2$         & $3.0\sim3.4$           & $4.22\pm0.10$     & 45.7  \\
                        & $J_{4\mu}$         &  $4.5^2$         & $3.0\sim3.3$           & $4.07\pm0.10$     & 44.4
\vspace{5pt} \\
                        & $J_{3\mu}$         &  $10.9^2$        & $8.5\sim9.5$           & $10.32\pm0.09$    & 47.0 \\
$qb\bar q\bar b$ system & $J_{4\mu}$         &  $10.8^2$        & $8.5\sim9.2$           & $10.22\pm0.11$    & 44.6 \\
                        & $J_{7\mu}$         &  $10.7^2$        & $7.8\sim8.4$           & $10.14\pm0.10$    & 44.8  \\
                        & $J_{8\mu}$         &  $10.7^2$        & $7.8\sim8.4$           & $10.14\pm0.09$    & 44.8
\vspace{5pt} \\
                        & $J_{3\mu}$         &  $10.9^2$        & $8.5\sim9.5$           & $10.34\pm0.09$    & 46.1 \\
$sb\bar s\bar b$ system & $J_{4\mu}$         &  $10.8^2$        & $8.5\sim9.1$           & $10.25\pm0.10$    & 43.3 \\
                        & $J_{7\mu}$         &  $10.8^2$        & $7.5\sim8.6$           & $10.24\pm0.11$    & 47.1  \\
                        & $J_{8\mu}$         &  $10.8^2$        & $7.5\sim8.6$           & $10.24\pm0.10$    & 47.1  \\
\hlinewd{.8pt}
\end{tabular}
\tabcaption{The threshold value, Borel window, mass and pole
contribution corresponding to the currents with $J^{PC}=1^{++}$ in the
$qc\bar q\bar c$, $sc\bar s\bar c$, $qb\bar q\bar b$ and $sb\bar
s\bar b$ systems.\label{table4}}
\end{center}
\begin{center}
\begin{tabular}{cccccc}
\hlinewd{.8pt}
                   & Currents & $s_0(\mbox{GeV}^2)$&$[M^2_{\mbox{min}}$,$M^2_{\mbox{max}}](\mbox{GeV}^2)$&$m_X$\mbox{(GeV)}&PC(\%)\\
\hline
                        & $J_{3\mu}$         &  $4.6^2$            & $3.0\sim3.4$           & $4.16\pm0.10$     & 46.2  \\
$qc\bar q\bar c$ system & $J_{4\mu}$         &  $4.5^2$            & $3.0\sim3.3$           & $4.02\pm0.09$     & 44.6  \\
                        & $J_{5\mu}$         &  $4.5^2$            & $3.0\sim3.4$           & $4.00\pm0.11$     & 46.0  \\
                        & $J_{6\mu}$         &  $4.6^2$            & $3.0\sim3.4$           & $4.14\pm0.09$     & 47.0
\vspace{5pt} \\
                        & $J_{3\mu}$         &  $4.7^2$            & $3.0\sim3.6$           & $4.24\pm0.10$     & 49.6  \\
$sc\bar s\bar c$ system & $J_{4\mu}$         &  $4.6^2$            & $3.0\sim3.5$           & $4.12\pm0.11$     & 47.3  \\
                        & $J_{5\mu}$         &  $4.5^2$            & $3.0\sim3.3$           & $4.03\pm0.11$     & 44.2  \\
                        & $J_{6\mu}$         &  $4.6^2$            & $3.0\sim3.4$           & $4.16\pm0.11$     & 46.0
\vspace{5pt} \\
                        & $J_{3\mu}$         &  $10.6^2$           & $7.5\sim8.5$           & $10.08\pm0.10$    & 45.9  \\
$qb\bar q\bar b$ system & $J_{4\mu}$         &  $10.6^2$           & $7.5\sim8.5$           & $10.07\pm0.10$    & 46.2  \\
                        & $J_{5\mu}$         &  $10.6^2$           & $7.5\sim8.4$           & $10.05\pm0.10$    & 45.3  \\
                        & $J_{6\mu}$         &  $10.7^2$           & $7.5\sim8.7$           & $10.15\pm0.10$    & 47.6
\vspace{5pt} \\
                        & $J_{3\mu}$         &  $10.6^2$           & $7.5\sim8.3$           & $10.11\pm0.10$    & 43.8  \\
$sb\bar s\bar b$ system & $J_{4\mu}$         &  $10.6^2$           & $7.5\sim8.4$           & $10.10\pm0.10$    & 44.1  \\
                        & $J_{5\mu}$         &  $10.6^2$           & $7.5\sim8.3$           & $10.08\pm0.10$    & 43.7  \\
                        & $J_{6\mu}$         &  $10.7^2$           & $7.5\sim8.5$           & $10.18\pm0.10$    & 46.5  \\
\hlinewd{.8pt}
\end{tabular}
\tabcaption{The threshold value, Borel window, mass and pole
contribution corresponding to the currents with $J^{PC}=1^{+-}$ in the
$qc\bar q\bar c$, $sc\bar s\bar c$, $qb\bar q\bar b$ and $sb\bar
s\bar b$ systems.\label{table5}}
\end{center}

\section{Conclusion}\label{sec:conclusion}

We have performed the QCD sum rule analysis with tetraquark
charmonium-like currents in vector and axial-vector channels.
The two-point correlation functions and the spectral densities
for all the interpolating currents have been evaluated at the
quark-hadron level. The numerical analysis shows that the four
quark condensate $\qq^2$ is the dominant power contribution to
the OPE series for all the vector channel currents. In the situation
of the axial-vector channel currents, however, the most important
corrections are the quark condensate $\qq$. These
properties of the spectral densities lead to a better OPE
convergence for the currents in the vector channel than that in
the axial-vector channel. The $m_s$ related terms in the OPE series
of $sc\bar s\bar c$ systems lead to more stable mass sum rules than
that of the $qc\bar q\bar c$ systems.
In the working range of the Borel parameter, only the currents
$J_{1\mu}, J_{4\mu}$ and $J_{7\mu}$ with $J^{PC}=1^{--}$ display
stable QCD sum rules in the $qc\bar q\bar c$ system. The extracted
mass is around $4.6\sim 4.7$ GeV from these currents, which is
consistent with the mass of the meson $Y(4660)$. This result
implies a possible tetraquark interpretation for $Y(4660)$. In the
$sc\bar s\bar c$ system, all currents except $J_{5\mu}, J_{6\mu}$
have stable QCD sum rules and the extracted mass is about $4.6\sim 4.9$ GeV.
The Borel window for the currents in the
axial-vector channel is very small because of the bad OPE
convergence. For the currents with $J^{PC}=1^{++}$ in the $qc\bar
q\bar c$ system, only $J_{3\mu}$ and $J_{4\mu}$ have reliable QCD
sum rules. The same situation occurs in the $sc\bar s\bar c$
system. The extracted masses are about $4.0\sim 4.2$ GeV, which is
$0.1\sim 0.3$ GeV higher than the mass of $X(3872)$.

The possible decay modes of these charmonium-like states are also studied
by considering the conservation of the angular momentum, P-parity, C-parity,
isospin and G-parity~\cite{2011-Chen-p34010-34010}.

\section*{Acknowledgments}
The authors thank Professor W. Z. Deng for useful discussions.
This project was supported by the National Natural Science
Foundation of China under Grants 10625521,10721063 and Ministry of
Science and Technology of China(2009CB825200).


\end{document}